\documentclass[12pt]{iopart}
\usepackage{natbib}
\usepackage{lscape}
\usepackage{rotating}
\usepackage{overpic}
\usepackage{longtable}
\usepackage{amssymb}
\begin{document}

\title[Nova Light Curves with SMEI]{Exquisite Nova Light Curves from the Solar Mass Ejection Imager (SMEI)}

\author{R Hounsell$^1$, M F Bode$^1$, P P Hick$^2$, A Buffington$^2$, B V Jackson$^2$, J M Clover$^2$, A W Shafter$^3$, M J Darnley$^1$, N R Mawson$^1$, I A Steele$^1$, A Evans$^4$, S P S Eyres$^5$ \& T J O'Brien$^6$}

\address{$^1$Astrophysics Research Institute, Liverpool John Moores University, Birkenhead, CH41~1LD, UK}

\address{$^2$Center for Astrophysics and Space Sciences, University of California, San Diego, 9500 Gilman Drive 0424, La Jolla, CA 92093-0424, USA}

\address{$^3$Department of Astronomy, San Diego State University, San Diego, CA 92182, USA}

\address{$^4$Astrophysics Group, Keele University, Keele, Staffordshire, ST5~5BG, UK}

\address{$^5$Jeremiah Horrocks Institute, University of Central Lancashire, Preston, PR1~2HE, UK}

\address{$^6$Jodrell Bank Centre for Astrophysics, School of Physics and Astronomy, University of Manchester, Manchester, M13~9PL, UK}

\ead{rah@astro.livjm.ac.uk}

\begin{abstract}
We present light curves of three classical novae (KT Eridani, V598 Puppis, V1280 Scorpii) and one recurrent nova (RS Ophiuchi) derived from data obtained by the Solar Mass Ejection Imager (SMEI) on board the Coriolis satellite. SMEI provides near complete sky-map coverage with precision visible-light photometry at 102-minute cadence. The light curves derived from these sky maps offer unprecedented temporal resolution around, and especially before, maximum light, a phase of the nova eruption normally not covered by ground-based observations. They allow us to explore fundamental parameters of individual objects including the epoch of the initial explosion, the reality and duration of any pre-maximum halt (found in all three fast novae in our sample), the presence of secondary maxima, speed of decline of the initial light curve, plus precise timing of the onset of dust formation (in V1280 Sco) leading to estimation  of the bolometric luminosity, white dwarf mass and object distance. For KT Eri, Liverpool Telescope SkyCamT data confirm important features of the SMEI light curve and overall our results add weight to the proposed similarities of this object to recurrent rather than to classical novae. In RS Oph, comparison with hard X-ray data from the 2006 outburst implies that the onset of the outburst coincides with extensive high velocity mass-loss. It is also noted that two of the four novae we have detected (V598 Pup and KT Eri) were only discovered by ground-based observers weeks or months after maximum light, yet these novae reached peak magnitudes of 3.46 and 5.42 respectively. This emphasizes the fact that many bright novae per year are still overlooked, particularly those of the very fast speed class. Coupled with its ability to observe novae in detail even when relatively close to the Sun in the sky, we estimate that as many as 5 novae per year may be detectable by SMEI.
\end{abstract}

\pacs{90}
\ams{85-05}
\submitto{The Astrophysical Journal}
\noindent{\it Keywords\/}: Cataclysmic Variables, transients, binary stars, novae, space-borne photometry

%\maketitle

\section{Introduction}
\label{sec:intro}
Classical novae (CNe) are all close binary systems that form a subclass of the cataclysmic variables \citep[see][2008 for reviews]{war95}. In these systems a late type star (usually a Main Sequence dwarf) fills its Roche lobe and transfers mass to a white dwarf (WD) companion via an accretion disk. The outbursts of classical novae (and recurrent novae, RNe) are caused by a thermonuclear runaway (TNR) in the material accreted onto the surface of the WD \citep[see][for reviews]{Starrfieldbook, 1989PASP..101....5S}. The resulting energy release ($\sim$~$10^{44}$ - $10^{45}$ ergs) is sufficient to expel the accreted envelope and drive substantial mass loss (10$^{-5}$ - 10$^{-4}$~M$_{\odot}$ in CNe, perhaps two orders of magnitude less than this in RNe) from the system at high velocities (of order a few hundred to several thousand km s$^{-1}$).

Based on an extrapolation of the observed nova density in the solar neighborhood, \citet{1997ApJ...487..226S} has estimated a Galactic nova rate of approximately 35~yr$^{-1}$. Of these an average of roughly one CN per year has been {\it observed} to reach $m_{V}$ = 8 or brighter \citep[see Figure 2 of][]{2002AIPC..637..462S}. Since these historical observations are clearly incomplete at $m_{V}$ = 8, the actual number of novae reaching this brightness is expected to be significantly higher \citep[see also][]{war08}.

The Solar Mass Ejection Imager (SMEI) is housed on board the Coriolis satellite which was launched on 2003 January 6th into a 840 km Sun-synchronous terminator orbit \citep{2003SoPh..217..319E}. The instrument consists of three baffled CCD cameras each with a 60$^{\circ}$$\times$3$^{\circ}$ field of view, combining to sweep out a 160$^{\circ}$ arc of sky \citep{2007SPIE.6689E...9H}. Peak quantum efficiency of the instrument is at approximately 700 nm with a FWHM $\sim$ 300 nm. SMEI maps out nearly the entire sky with each orbit of the spacecraft (102 minutes). SMEI is operated as a high precision differential photometer \citep{2006ApJ...637..880B, AB} and can reliably detect brightness changes in point sources down to $m_{\mathrm{SMEI}} \sim$ 8 (see Section \ref{sec:data} for details on $m_{\mathrm{SMEI}}$). Therefore one class of optical variables that are potentially within the detection limit of SMEI is CNe. The instrument is specifically designed to map large-scale variations in heliospheric electron densities from Earth orbit by observing the Thomson-scattered sunlight from solar wind electrons in the heliosphere \citep{2004SoPh..225..177J}. In order to isolate the faint Thomson-scattered sunlight the much larger white-light contributions from the zodiacal dust cloud, the sidereal background and individual point sources (bright stars and planets) must be determined and removed. Thus, brightness determination of point sources is a routine step in the SMEI data analysis, capable of providing stellar time series with a 102-minute time resolution.

The results of \citet{2002AIPC..637..462S} indicate that as many as $\sim$ 5 CNe per year are potentially detectable by SMEI. The high cadence of SMEI along with its ability to monitor objects appearing closer to the Sun than is possible from ground-based observations, makes it feasible not only to constrain the observed nova rate, but also to measure nova light curves near and especially before maximum light with unprecedented temporal resolution.

In this {\it paper} we present light curves of four bright CNe ($m_{V}^{\rm max} < 8$) that were detected by SMEI over the past seven years.

\section{SMEI data acquisition and reduction}
\label{sec:data}

The SMEI cameras continuously record CCD image frames with an exposure time of 4 seconds. Scanning nearly the entire sky in each 102-minute orbit, approximately 1500 frames per camera per orbit are available to compose a full-sky map. The processing steps used at UCSD to convert the raw CCD images into photometrically accurate white-light sky maps include: integration of new data into the SMEI database; removal of an electronic offset (bias) and dark current pattern; identification of cosmic rays, space debris and ``flipper pixels'' \citep[see][for further details]{2005SPIE.5901..340H}; and placement of the images onto a high-resolution sidereal grid using spacecraft pointing information. The CCD pixel size is about 0.05$^{\circ}$. Due to telemetry considerations the CCD frames are re-binned on board by 2 x 2 (camera 3) or 4 x 4 pixels (cameras 1 and 2), resulting in a ``science mode'' resolution of 0.1$^{\circ}$ or 0.2$^{\circ}$. The CCD frames are assembled into a skymap using a fixed sidereal grid with a resolution of 0.1$^{\circ}$ in Right Ascension and declination, commensurate with the science mode resolution of the CCD frames. As the instrument orbits the Earth the 3$^{\circ}$ narrow dimension of the cameras sweeps across the sky. A specific sky location is inside the field of view for typically a minute or more (depending on camera and sky location). With a 4 second exposure this implies that about a dozen or more separate measurements from sequential CCD frames are available for each sidereal skybin. These are combined to provide one measurement per orbit. Point sources, including novae, are fit from these sidereal maps.

%Moving through its 102-minute Sun-synchronous polar orbit the SMEI cameras continuously take 4 second exposures, resulting in $\simeq$ 1500 frames per camera per orbit. The data from SMEI are collected via down-links to several ground stations. Raw CCD images are then processed to create photometrically accurate white-light sky maps. The University of California, San Diego (UCSD) data processing steps include: integration of new data into the SMEI database; removal of an electronic offset (bias) and dark current pattern; identification of cosmic rays, space debris and ``flipper pixels'' \citep[see][for further details]{2005SPIE.5901..340H}; and placement of the images onto a high-resolution sidereal grid using spacecraft pointing information. \textbf{Each CCD pixel in this grid has a resolution of $\sim 0.05^{\circ}$. The grid is then converted to a lower-resolution set of sidereal maps of sky brightness with a resolution of $\sim 0.1^{\circ}$}. It is from these maps that point sources including novae are fit. \textbf{As the instrument orbits the Earth, the 3$^{\circ}$ narrow dimension of the camera sweeps past a given sky location in about 1 minute. Therefore about a dozen sequential CCD frames contribute to the brightness of a point source.}%

The standard point spread function (PSF) used for fitting was created via the observation of several bright isolated stars over a year \citep[see][]{2007SPIE.6689E...9H}. The resulting PSF shape has a full width of approximately 1$^{\circ}$, is highly asymmetric, and ``fish-like'' in appearance; this is caused by comatic and spherical aberrations of the optics, and varies somewhat with pixel position in each of the three cameras \citep[see Figure 1 of][]{2007SPIE.6689E...9H}. 

The UCSD SMEI database contains a list (location in the sky and apparent magnitude) of 5600 sources expected to be brighter than 6$^{th}$ magnitude in the SMEI skymaps (i.e. $m_{\mathrm{SMEI}}<6$). These sources (and the brightest planets) are fitted to the standard PSF using a least-squares fitting procedure implemented in IDL (Interactive Data Language). In its simplest form the fit provides an analytic solution for a planar background and the brightness of the point source under examination. At the expense of a substantial increase in computational resources the quality of the fit can be improved by also iteratively fitting the PSF centroid, PSF width, and PSF orientation \citep[see][for further details]{2005SPIE.5901..340H, 2007SPIE.6689E...9H}.

Star crowding occasionally requires that multiple stars are fitted simultaneously. A star of interest is considered crowded when it lies less than one PSF width from another bright star (typically 6$^{th}$ magnitude or brighter). Currently the simultaneous fitting of stars is conducted when stellar separation is within $0.25^{\circ}-0.75^{\circ}$. In this instance the brightness contributions of the two stars can be separated and contamination of the sources is considered minimal. However, if stellar separation is less than 0.25$^{\circ}$ the stars can not be separated. The brighter star is fit first and is assumed to include the brightness of the fainter star (which is not fit at all), the source is then considered to be contaminated. We are currently able to fit four bright stars simultaneously using the least-squares fit, therefore many crowded regions of the sky such as close to the Galactic plane are off-limits\footnotemark[1]. The surrounding region of each point of interest must therefore be assessed on an object-by-object basis for levels of potential contamination.

\footnotetext[1]{The lowest Galactic latitude where a nova has been detected without contamination is 1.5$^{\circ}$}

For the purpose of this paper a supplementary star catalogue was added to the UCSD SMEI data base. This catalogue contains the names, co-ordinates, and discovery magnitudes of 57 CNe and 2 RNe (RS Oph and U Scorpii) with eruptions dating between 2003 and 2010. As an initial trial, 22 of the brightest novae were examined by visually inspecting the composite skymaps produced by the SMEI data pipeline using the tools provided in the SMEI data analysis software. In total 13 erupting novae were detected and investigated in further detail. Photometry of each nova was obtained using the extended least-squares fit described above (i.e. including the iterative fitting of PSF centroid, width, and orientation). Zodiacal and sidereal background light were also taken into account at this stage. Prior to fitting, the surrounding area of each nova was examined for stellar contamination caused by star crowding and where possible a simultaneous fit was conducted in order to obtain the most accurate photometry. It was found that only one of the novae presented here (V598 Pup, see Section~\ref{sec:v598Pup}) possessed a bright neighboring star. The fitted value for the flux of the nova (and neighboring star where applicable) was then converted into an unfiltered SMEI apparent magnitude ($m_{\mathrm{SMEI}}$).

The remaining 9 novae were not clearly observed due to several effects. Many of these novae have low (mag$\sim 7-8$) peak magnitudes making detection difficult, some were missed due to technical difficulties with the imager itself, or due to transient stray light from the Sun and/or planets \citep[e.g. the U Sco 2010 outburst,][was missed due to its location within the 20$^{\circ}$ mask of the Sun]{sch10}, finally a few novae were located in such densely populated regions that obtaining accurate photometry of the object would be impossible. 
 
Four of the 13 novae detected produced highly detailed nova light curves. Each of these four light curves has been examined for extra sources of noise (e.g. space debris in frames, stray light from crossing planets, etc.) and the validity of the fit checked. These four uniquely detailed light curves are discussed here.

\section{Light curves}
\label{sec:lightcures}
The light curves presented in this section are unprecedented in their detail and present data on phases that previously were both poorly covered observationally, and are poorly understood. Table~\ref{table1} summarises our main findings.

\subsection{RS Ophiuchi}
\label{sub:RSOph}
RS Ophiuchi is a recurrent nova whose latest outburst was first observed by \citet{nar06} on 2006 February 12.83 UT (MJD 53778.83) at $m_{V} = 4.5$. The 2006 outburst was observed in great detail across the electromagnetic spectrum \citep[see][and references therein]{eva08}. Of particular note was the interaction of the high velocity ejecta with the pre-existing wind of the red giant in the system, leading to the rapid establishment of strong shocks \citep[e.g.][]{bod06}. 

\begin{figure}[h]
\centering
\noindent\makebox[\textwidth]{%
\includegraphics[scale=0.8]{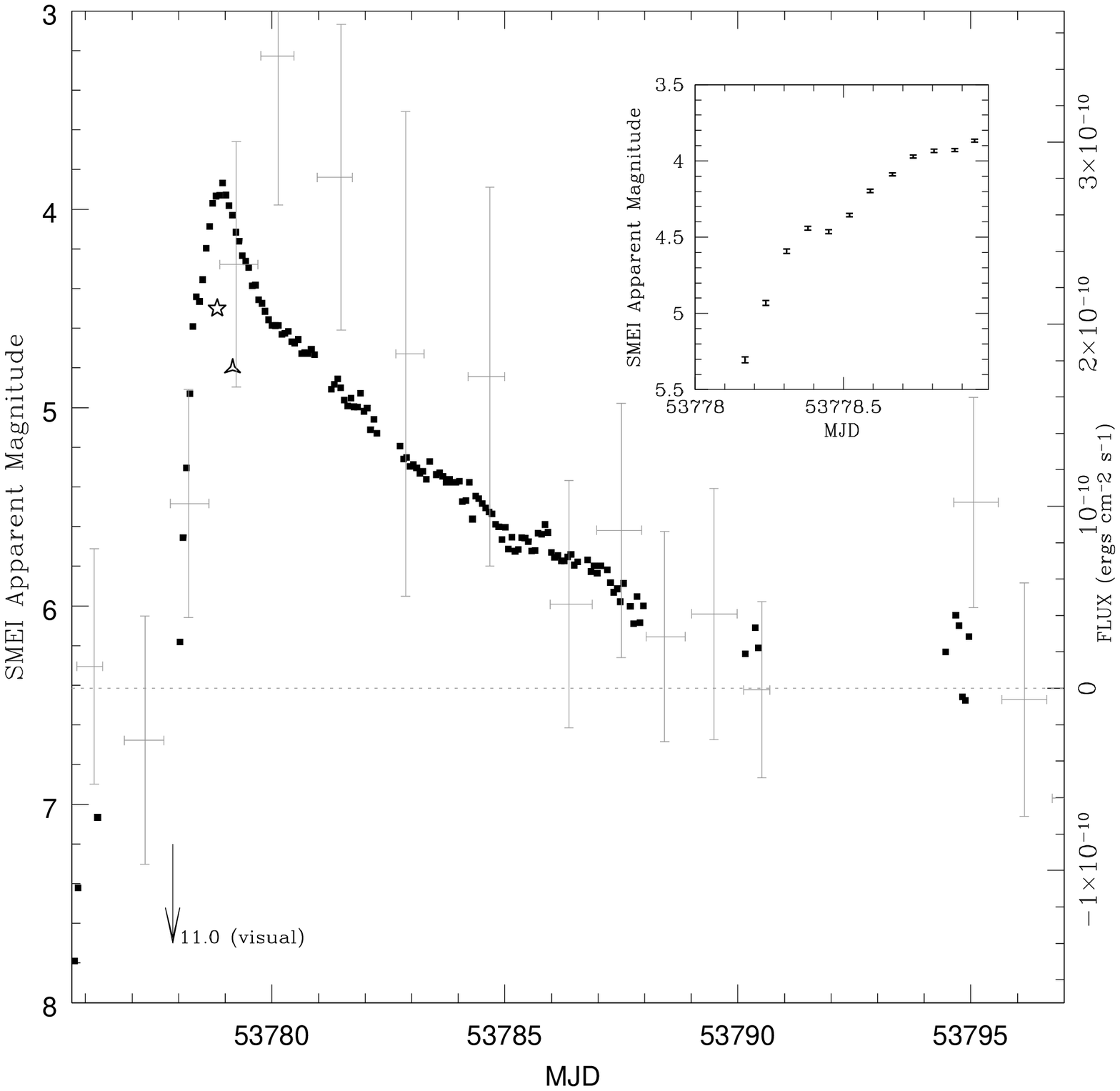}}
\caption{SMEI light curve of RS Oph  (black squares) in terms of ``SMEI magnitude'' \citep[see][]{AB} versus time (left-hand y-axis). Over plotted (gray) are the \textit{Swift} BAT 14-25 keV data from \cite[][right-hand y-axis]{bod06}, the gray dashed line indicates zero flux on the right-hand y-axis. The star represents the discovery magnitude of the nova taken from \cite{nar06}. The triangle is the peak magnitude listed by the AAVSO\footnotemark[3]. The apparent discrepancy between ground-based and SMEI magnitudes is discussed in Section~\ref{sub:RSOph}. An arrow is used to indicate the latest observed magnitude of the nova before rise, according to the AFOEV\footnotemark[4] data set. The inset shows the rising portion of the light curve with an expanded time scale.}
\label{figure1}
\end{figure}

RS Oph showed a very rapid rise to maximum in the SMEI data (see Figure~\ref{figure1}), increasing in brightness by 2.3 magnitudes in 0.9 days. This value is measured using the first reliable detection of the nova on its rise to maximum and the peak magnitude itself. The SMEI light curve shows clear evidence of a pre-maximum halt (see inset in Figure~\ref{figure1}) starting 2006 February 12.31 UT (MJD 53778.31) and lasting just a few hours with a mean $m_{\mathrm{SMEI}} =  4.50 \pm 0.05$ (defined as the mean magnitude over the duration of the halt, quoted error is the RMS scatter). The duration of the halt (for RS Oph, V598 Pup and KT Eri) is taken to be the time between the first and third change in the gradient of the rising light curve and is appropriate for the speed of the nova as proposed by e.g. \citet{pay64}. Note that this halt (and subsequent ones in other novae - see below) looks like a temporary reversal in the light curve and may be related to a change in mass loss rate but this needs to be investigated in detail. Peak brightness of the nova was reached on 2006 February 12.94 (MJD 53778.94) $\pm 0.04$ UT at $m_{\mathrm{SMEI}} = 3.87 \pm 0.01$ (we note that the peak magnitude derived from the SMEI data is over half a magnitude brighter than ground-based estimates. This discrepancy may be caused by a slight over subtraction within the fit of the SMEI PSF or the fact that the ground-based observations are visual magnitude estimates, compared to the broader band of SMEI, or both). Thereafter, the nova declined very rapidly with $t_2 = 7.9$ days \citep[$t_{n}$ is defined as the number of days it takes the nova to decline \textit{n} magnitudes from peak;][]{pay64}.

RS Oph remains the only nova to be detected at outburst with the \textit{Swift} Burst Alert Telescope \citep[BAT;][]{sen08,bod06}. It was clearly detected in the 14-25 keV channel for $\sim 5$ days around discovery, with a marginal detection in the 25-50 keV band at this time. Figure~\ref{figure1} shows the BAT 14-25 keV results over-plotted on the SMEI data.  It is apparent that the initial rise of the optical and hard X-ray is coincident within the temporal uncertainty. As the hard X-ray emission is thought to arise from the interaction of the fastest moving ejecta with the pre-outburst wind of the red giant \citep{bod06}, the coincidence of the onset of the outburst as seen in the optical with that found in the BAT data implies that significant high velocity mass loss occurs very early in the outburst itself. As the optical peak may indicate the time of highest mass loss rate from the surface of the WD, one might reasonably expect $[t_{\rm{max}}]_{\rm{BAT}} \gtrsim [t_{\rm{max}}]_{\rm{SMEI}}$ as appears to be the case here.

\subsection{V1280 Scorpii}
\label{sub:V1280Sco}
\begin{figure}[h]
\centering
\noindent\makebox[\textwidth]{%
\includegraphics[scale=0.8]{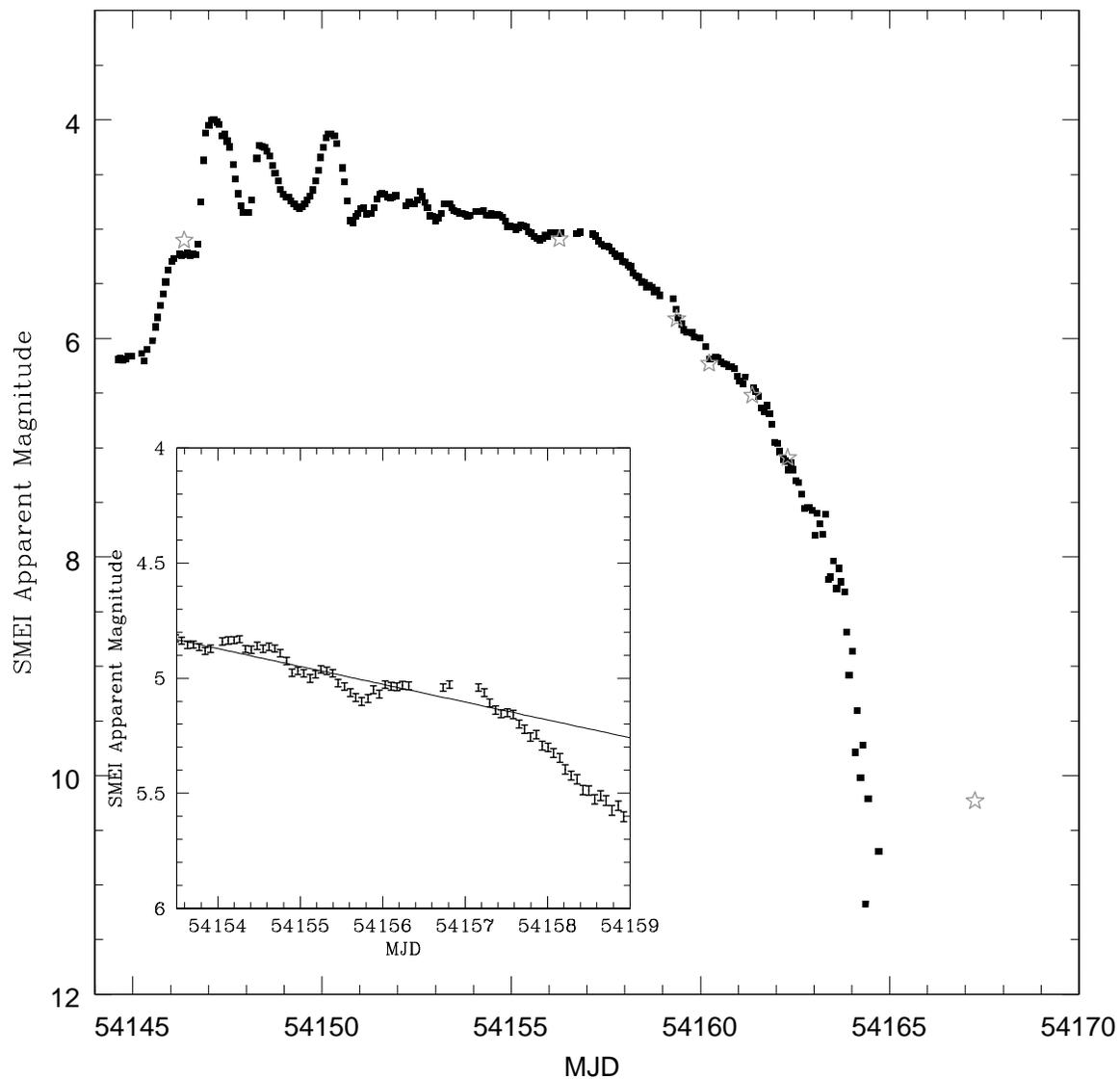}}
\caption{SMEI light curve of V1280 Sco (black squares), superimposed (gray stars) are data from the ``$\pi$ of the Sky'' project. The inset shows the region around the light curve break which is associated with the onset of dust formation. The solid line shows the fit to the pre-break SMEI light curve and its extrapolation.}
\label{figure2}
\end{figure}

V1280 Sco was discovered in outburst by \citet{2007CBET..834....1Y} on 2007 February 4.86 UT (MJD 54135.86). Twelve days later it reached visual maximum quoted as $m_{V} = 3.79$ \citep{2007CBET..852....1M}. Although the initial rise of the nova is lost in the SMEI data, due to transient stray light from the Sun, and from Jupiter as it moves across the sky, Figure~\ref{figure2} shows that the climb to maximum on 2007 February 16 is very slow (consistent with 12 days), as has been noted by various authors \citep[e.g.][]{2008A&A...487..223C}. 

%SMEI data indicate a pre-maximum halt occurring at February 14.?? UT with $m_{SMEI}$ $\sim$ 5.8 $\pm$ ??. This pre-maximum halt lasts for approximately 17 hours ending on February 15.?? UT. 
Canonically, the supposed pre-maximum halt is defined as occurring one to two magnitudes below peak optical brightness \citep[see][and references therein]{war08}. With this in mind there appears to be a halt before the first maximum of V1280 Sco lasting 0.42 days (duration of halt is defined here as the time between the first and second change in gradient of the rising light curve) with a mean $m_{\mathrm{SMEI}} = 5.231 \pm 0.003$ (see inset in Figure~\ref{figure2}). However, we note that there is evidence of an earlier plateau in the nova light curve, but which is not within the magnitude range expected.
Peak visual magnitude occurred on 2007 February 16.15 (MJD 54147.15) $\pm 0.04$ UT with $m_{\mathrm{SMEI}} = 4.00 \pm 0.01$ (see Figure~\ref{figure2}). The nova then experiences two major episodes of re-brightening peaking at February 17.34 (54148.34 MJD) $\pm$ 0.04 UT and 19.18 (MJD 54150.18) $\pm 0.04$ UT, with $m_{\mathrm{SMEI}} = 4.23 \pm 0.01$ and $4.13 \pm 0.01$ respectively. The existing published visual light curves lack such fine detail \citep[see e.g.][Figure 1]{das08}. Data from the ``$\pi$ of the Sky''\footnotemark[2] project are superimposed in Figure~\ref{figure2}. These are white light unfiltered magnitudes, confirming the SMEI calibration and following the general trend of the light curve. These data contain the best known pre-maximum values for the nova from a homogeneous observational set and illustrate our current lack of coverage of this phase of evolution. The subsequent decay of the SMEI light curve is marked by a distinct change in decline rate in visual light on 2007 February 26.4 (MJD 54157.37) $\pm 0.1$ UT (see inset in Figure~\ref{figure2}) at $m_{\mathrm{SMEI}} = 5.14 \pm 0.02$. The overall decline that then ensues is thought to be the effect of rapid formation of dust in the nova ejecta \citep{2007AAS...211.5116R,das08}.

\footnotetext[2]{http://grb.fuw.edu.pl/pi/index.html}

We may identify the change in slope on February 26.4 UT (54157.37 MJD) with the onset of large-scale dust formation in the ejecta. \citet{2008A&A...487..223C} note that the first unambiguous evidence of dust emission dominating the near-infrared spectra is on March 7, but they suggest that the absence of obvious emission in the spectrum of February 26.97 UT (MJD 54157.97) does not rule out the presence even at that stage of an extended optically thin dust shell. Certainly, the change in light curve slope on February 26.4 UT (MJD 54157.37) is a subtle effect that can only be derived from photometry with the temporal sampling and small intrinsic scatter of the SMEI data.

As noted, the rise to maximum light was very slow. From consideration of infrared photometry of the fireball expansion phase, \citet{das08} find that the outburst commenced $\sim 2.35$ days before discovery, on 2007 February 2.5 UT (MJD 54133.5). Assuming that extensive mass loss began at this time, from our SMEI results, this gives the condensation time of dust grains from the ejecta as $t_{c} \sim 24$ days. This timescale, together with the observed ejection velocity \citep[$\sim 600$ km s${^{-1}}$,][]{das08} and an assumed condensation temperature of dust grains \citep[$T_c$ = 1200K,][]{geh08,er08} leads to an estimate of the nova's luminosity at this time, assuming the nucleation centers act as black bodies as

\vspace{3mm}

$L_\star = 2.4\times10^4~({t_c}/{24~\rm{days}})^2~({v_{ej}}/{600~\rm{km~s}^{-1}})^2~({T_{\rm c}}/1200K)^4~{\rm L_\odot}$

\vspace{3mm}

Taking this as the Eddington luminosity of the WD \citep[e.g.][]{geh08} in turn implies $M_{\rm WD} = 0.6~$M$_\odot$. We note that the equilibrium temperature of the nucleation centers may be higher than that of a black body for the same $L_\star$ and distance from the nova, hence $M_{\rm WD}$ is likely to be an upper limit. This compares with the $M_{\rm WD}$ = 1 to 1.25~$M_\odot$ estimated by \cite{das08} from consideration of the timescale of mass loss, plus outburst amplitude $A$, and expansion velocity $v_{\rm exp}$. These authors admit however that such a high mass estimate is incompatible with what appears spectroscopically to be an explosion on a carbon-oxygen WD, for which our estimate of $M_{\rm WD}$ would be compatible. 

The derived $L_\star$ and a spectrum near maximum light akin to that of an F star (Bolometric Correction $\sim 0$) gives $M_V$ = -6.2. Taking the line-of-sight (interstellar) extinction to be $A_V = 1.2 \pm0.3$ \citep{2008A&A...487..223C} and $m_{V}^{\rm max} = 4$ yields a distance to the nova of $d = 630 \pm 100$ pc, roughly half that derived by \citet{2008A&A...487..223C}. We use a linear extrapolation of the nova light curve between 2007 February 20.59 (MJD 54151.59) and 26.59 UT (MJD 54157.59) in order to determine $t_{3}$. The data for the extrapolation are taken after the last re-brightening event, but before the dust break (i.e. removing the influence of dust formation) shown in Figure~\ref{figure2}. A $t_{3}$ value of $\sim 34$ days is determined, i.e. an estimated decline of 0.1 magnitude per day. We note that from the Maximum Magnitude-Rate of Decline relation (MMRD) given in \citet{2000AJ....120.2007D}, $M_{V} \sim -8$. However, the applicability of the MMRD is questionable in the context of such gross variability around maximum light, followed by a slow and steady decline.

\subsection{V598 Puppis}
\label{sec:v598Pup}
V598 Pup was discovered by \citet{2007ATel.1282....1R} in the XMM-Newton slew survey on 2007 October 9.0 UT (MJD 54382) as a transient X-ray source, designated XMMSLI J070542.7-381442. It was later identified as a nova by \citet{2007ATel.1285....1T} whilst trying to identify the object's optical counterpart. The peak visual magnitude was noted by \citet{2007IAUC.8899....2P} as $m_{V} \leq 4$ on 2007 June 5.968 UT (MJD 54256.968).

\begin{figure}[h]
\centering
\noindent\makebox[\textwidth]{%
\includegraphics[scale=0.8]{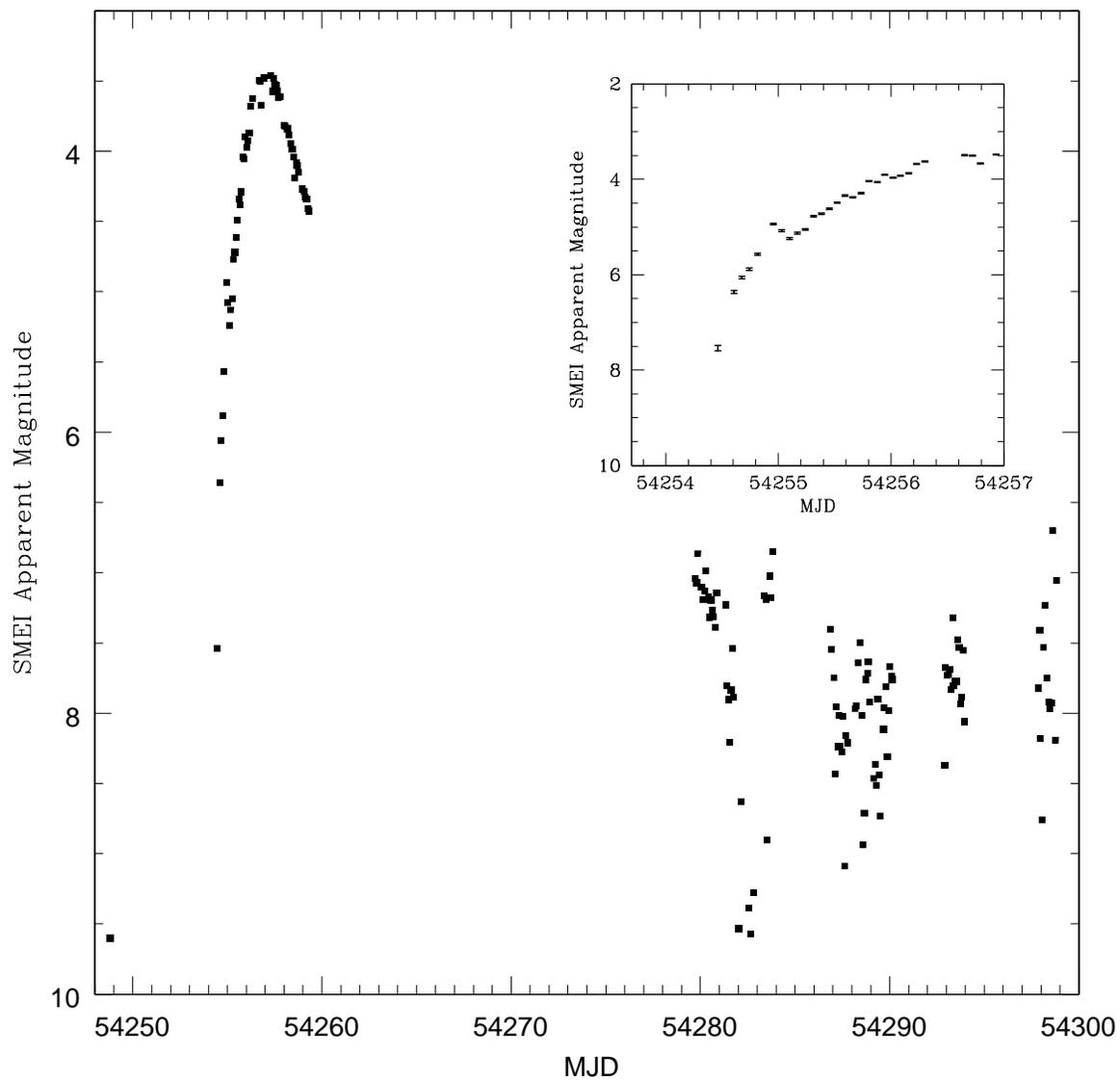}}
\caption{SMEI light curve of V598 Pup. The inset shows the rising portion of the light curve with an expanded time scale. Note that the variation of the light curve at MJD $\gtrsim$ 54280 is most likely due to contamination from a bright neighbouring star and problems within the fitting procedure, see Section~\ref{sec:v598Pup} for further details.}
\label{figure3}
\end{figure}

From the SMEI data shown in Figure~\ref{figure3} we find the rise to maximum to be very steep with the nova increasing 4.1 magnitudes within 2.8 days. A pre-maximum halt is indicated on 2007 June 3.82 UT (MJD 54254.82) with a mean $m_{\mathrm{SMEI}} = 5.2 \pm 0.1$ and duration a few hours (see inset in Figure~\ref{figure3}). The nova then rose to its peak visual magnitude of $m_{\mathrm{SMEI}} = 3.46 \pm 0.01$ on 2007 June 6.29 (MJD 54257.29) $\pm 0.04$ UT. Decline from maximum also appears steep, however a section of this decline phase has been missed in SMEI data due to a failure of the star tracker, causing the spacecraft to assume a Sun pointing mode. This failure lasted $\sim 21$ days. An estimate of $t_{2}$ using an extrapolated linear fit to the initial decline of the nova (between 2007 June 6.29 [MJD 54257.29] and 8.33 [MJD 54259.33] UT) yields $t_{2} = 4.3$ days.

It should be noted that V598 Pup is located close ($\sim 0.1^{\circ}$) to HD 54153, a 6$^{th}$ magnitude star. In order to reduce the star's effect on the nova a forced simultaneous fit was conducted. This procedure is ideally suitable for larger stellar separations (between $0.25^{\circ}-0.75^{\circ}$, see Section~\ref{sec:data} for further details) and thus can not consistently remove the contaminating star especially as the nova starts to fade. The variability seen in the light curve of Figure~\ref{figure3} at later times (MJD $\gtrsim 54280$) is therefore most likely due to contamination from the nearby bright star and problems occurring in the fitting procedure.

\subsection{KT Eridani}
\label{sec:KTEri}
KT Eri was discovered on 2009 November 25.536 UT (MJD 55160.536) by \citet{2009CBET.2050....1I} with an unfiltered CCD magnitude of 8.1. Like V598 Pup, KT Eri was missed at peak brightness and only discovered a considerable time later. Its outburst was found in pre-discovery images with a peak visual magnitude given as 5.4 on 2009 November 14.63 UT \citep[MJD 55149.63,][]{2009IAUC.9098....1Y}.
 
Pre- and post-outburst data for this object have been obtained by SMEI \citep{2010ATel.2558....1H} and the SkyCamT (SCT) instrument which is mounted to parallel-point with the main beam of the Liverpool Telescope (LT), La Palma \citep{ste04}. LTSCT uses a 35mm focal length lens to provide a $21^{\circ} \times 21^{\circ}$ field of view onto a $1024 \times 1024$ pixel detector, yielding a plate scale of 73.4 arcsec pixel$^{-1}$. The camera operates continuously throughout the night, taking a 10 second exposure once per minute in the direction of the main telescope pointing, giving a limiting magnitude of $\sim 12$. As with SMEI, the data are unfiltered (i.e. white light) and are calibrated with respect to four bright isolated USNO-B stars in the field of view. 

\begin{figure}[h]
\centering
\noindent\makebox[\textwidth]{%
\includegraphics[scale=0.8]{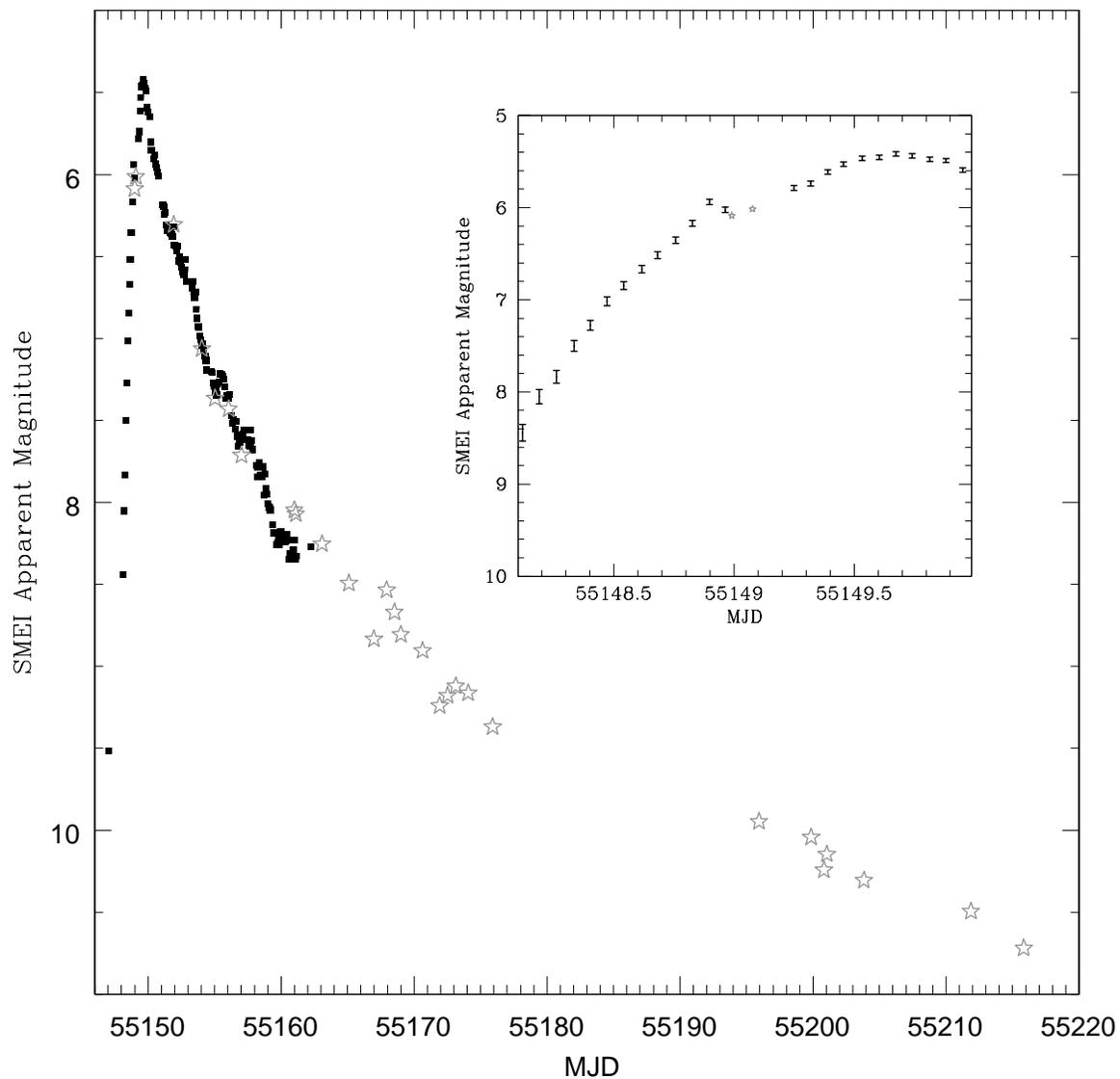}}
\caption{SMEI (black squares) light curve of KT Eri with Liverpool Telescope SkyCamT data superimposed (gray stars - see Section~\ref{sec:KTEri} for details). SMEI and LTSCT data seem to be in good agreement with each other confirming statements made within \citet{AB}. The inset shows the rising portion of the light curve with an expanded time scale}
\label{figure4}
\end{figure}

The SMEI and LT light curves are shown in Figure~\ref{figure4}. SMEI data indicate the initial rise of the nova is steep (rising 3.0 magnitudes over 1.6 days) first being clearly detected in outburst on 2009 November 13.12 UT (MJD 55148.12) with $m_{\mathrm{SMEI}} = 8.44 \pm 0.09$. Evidence of a pre-maximum halt occurring on 2009 November 13.83 (MJD 55148.83) $\pm 0.04$ UT with a mean $m_{\mathrm{SMEI}} = 6.04 \pm 0.07$ is given by SMEI with LTSCT observations adding two important points to the coverage of the halt (see inset in Figure~\ref{figure4}). The duration of this halt is again only a few hours. SMEI observations indicate that the nova reached maximum light on 2009 November 14.67 (MJD 55149.67) $\pm 0.04$ UT with $m_{\mathrm{SMEI}} = 5.42 \pm 0.02$. LTSCT observations bracket the peak seen with SMEI. The nova then subsequently declined rapidly with $t_2 = 6.6$ days confirming KT Eri as a very fast nova \citep{war08}. The last reliable SMEI detection of the nova occurred on 2009 November 27.23 (55162.23 MJD) $\pm 0.04$ UT at $m_{\mathrm{SMEI}} = 8.3 \pm 0.1$. LTSCT observations extend the optical coverage of the light curve untill 2010 January 19.85 UT (MJD 55215.85). The LTSCT data also confirm the calibration of the SMEI photometry and general trends in the resulting light curve. Similar results are also found within ``$\pi$ of the Sky'' data.

KT Eri has been detected as a radio source \citep{obr10} and a luminous soft X-ray source \citep{bod10}. Attention has been drawn to the similarities of its optical spectral and X-ray evolution to that of the recurrent nova LMC 2009a \citep{bod10}. Its outburst has also been associated with a highly variable stellar progenitor at mag $\sim 15$ showing evidence for pre-outburst circumstellar material and with similarities to the soft X-ray transient CSS081007:030559+054715 \citep{dra09}. We note that the very fast decline and relatively low amplitude of the outburst ($A \sim 10$ mags) place KT Eri in an anomalous position on the $A$ vs speed class diagram for CNe \citep[e.g.][]{war08}, but much more in line with that for recurrent novae such as U Sco \citep{sch10}.

\section{Discussion and Conclusions}
\label{sec:con}
This work has enabled us to follow in unprecedented detail the rise to maximum of all four of the novae surveyed. In turn, it has provided significant, detailed and undeniable evidence for the existence of the previously controversial pre-maximum halt, with accurate times of occurrence, duration and magnitude below peak given. The reality of this halt in all three of the fast novae observed (and possibly in a slightly different form in the slow nova V1280 Sco) is a challenge to detailed models of the nova outburst. From Table \ref{table1} it may also be noted that there does not seem to be a correlation between the properties of the pre-maximum halt ($\Delta m_{\mathrm{SMEI}}$, $\Delta t$, number of magnitudes below maximum, and time before peak) and the properties of the nova or its eruption (speed class), although our sample size is admittedly small at present.

The time of each nova's peak optical brightness has been derived with previously unobtainable accuracy, marking as it does the time of greatest extent of the pseudophotospheric radius in each object. Perhaps the most intriguing features around maximum light are displayed by V1280 Sco where two re-brightenings may be associated with epochs of enhanced mass loss from the WD surface. What the mechanism is that would lead to such enhancements during the TNR is a matter of conjecture. Within the initial decline of each nova light curve small oscillations can also be seen.

%Within the initial decline of each nova light curve small oscillations can be seen. This could indicate that there is no such thing as a smooth decline \citep{2010AJ....140...34S} however, more data are required to be definate here.
%Using the precise timing of the onset of dust formation of nova V1280 Sco we obtain a WD mass of  approximately 0.63 M$_\odot$. This value is representative of a CO WD and is in turn consistent with the spectra obtained by \citet{das08} rather than the larger WD mass those authors obtained.

Overall, this initial investigation of the SMEI data archive has proven how important it is to examine all-sky data with regards to transient events. As with the case of both novae V598 Pup and KT Eri, even the brightest (naked eye) novae may be missed by conventional ground-based observing techniques, \citet[][and references therein]{war89, war08} reached the same conclusion. \citet{2002AIPC..637..462S} estimates that as many as five novae with maxima brighter than 8$^{th}$ magnitude may occur each year, instead of the one or two typically observed. This estimate suggests that over 35 novae may be found within the SMEI data archive, 60-80$\%$ of which will be unknown. It is therefore the intention of the authors to continue the investigation of the SMEI archival data searching for previously known and unknown transient events from 2003 to the present day. Detection of previously unknown novae will be conducted with a SMEI algorithm designed to detect stellar variation within the data archive. 
Overall, the data provided by SMEI may well have opened a new chapter in our observations and understanding of novae.

%Overall, this initial investigation of the SMEI data archive has proven how important it is to examine all-sky data with regards to transient events. As with the case of both nova V598 Pup and KT Eri, even the brightest (naked eye) novae may be missed by conventional ground-based observing techniques. This point has been made by previous authors such as \citet{war89, war08} and \citet{2002AIPC..637..462S}, from these we note that five novae per year are expected to be brighter than eighth magnitude and thus detectable by SMEI. It is therefore the intention of the authors to continue the investigation of the SMEI archival data searching for transient events from 2003 to the present day. This investigation shall be conducted with IDL-formatted package \textit{smei\_findpoint}. 
%Overall, the data provided by SMEI may well have opened a new chapter in our observations and understanding of novae.  

\ack
The USAF/NASA SMEI is a joint project of the University of California San Diego, Boston College, the University of Birmingham (UK), and the Air Force Research Laboratory.
The Liverpool Telescope is operated on the island of La Palma by Liverpool John Moores University in the Spanish Observatorio del Roque de los Muchachos of the Instituto de Astrofisica de Canarias with financial support from the UK Science and Technology Facilities Council. We would like to thank Gerry Skinner for provision of the BAT data on RS Oph and pointing us to the AFOEV observations of the outburst and an anonymous referee for pertinent comments which helped improve the manuscript.
P. P. Hick, A. Buffington, B. V. Jackson, and J. M. Clover acknowledge support from NSF grant ATM-0852246 and NASA grant NNX08AJ11G.
A. W. Shafter acknowledges support from NSF grant AST-0607682.
R. Hounsell and N. R. Mawson acknowledge support from STFC postgraduate studentships.

\newpage
\begin{landscape}
\begin{center}
\begin{table}\footnotesize
\caption{\label{table1}Derived light curve parameters.}
\begin{tabular}{p{1.6cm} p{3.05cm} p{3.05cm} p{1.8cm} p{1cm} p{1cm} p{2.0cm} p{1.7cm} p{1.5cm} p{2.0cm}}
\br
Name & Onset of Outburst yyyy/mm/dd $\pm 0.04$ days & Time of Maximum yyyy/mm/dd $\pm 0.04$ days & Peak SMEI magnitude & $t_{2}$ (days) & $t_{3}$ (days) & Pre-max halt~duration (days)$^{\Psi}$ & Pre-max mean magnitude & $\Delta m_{\mathrm{SMEI}}$ from halt to peak$^{\phi}$ & $\Delta t$~from halt to peak (days)\\
\mr
RS Oph & 2006/02/12.03 & 2006/02/12.94 & 3.87 $\pm0.01$ & 7.9 & - & 0.14 & 4.50$\pm$0.05 & 0.63 & 0.49\\
V1280 Sco & - & 2007/02/16.15 & 4.00 $\pm0.01$ & 21.3$^\dagger$ & 34.3$^\dagger$ & 0.42 & 5.231$\pm$0.003 & 1.23 & 0.49\\
V598 Pup & 2007/06/3.47 & 2007/06/06.29 & 3.46 $\pm0.01$ & 4.3$^{\diamond}$ & - & 0.28 & 5.2$\pm$0.1 & 1.74 & 2.19\\
KT Eri & 2009/11/13.12 & 2009/11/14.67 & 5.42 $\pm0.02$ & 6.6 & 13.6$^{\star}$ & 0.14 & 6.04$\pm$0.07 & 0.63 & 0.71\\
\br
\end{tabular}
\end{table}
\end{center}
$^{\Psi}$The duration of the halt is taken to be the time between the first and third change in the gradient of the rising light curve for RS Oph, V598 Pup and KT Eri. With V1280 Sco it is taken to be the time between the first and second change in gradient of the rising light curve.\\
$^{\phi}\Delta$m from halt to peak is calculated using the mean magnitude of the pre-maximum halt\\
$^\dagger$Using an extrapolation ignoring dust extinction (see text for details)\\
$^{\diamond}$Using a linear extrapolation of the initial decline (see text for details)\\
$^{\star}$Using the extrapolation of SMEI and LT data.\\
\end{landscape}

\newpage

\footnotetext[3]{American Association of Variable Star Observers - http://mira.aavso.org/tmp/data5167.txt}
\footnotetext[4]{Association Francaise des Observateurs d'Etoiles Varuables - http://cdsarc.u-strasbg.fr/afoev/oph/rs}

\newpage

\newpage

\newpage

\clearpage
\def\newblock{\hskip .11em plus .33em minus .07em}

\end{document}